\documentclass[12pt]{article}

\normalbaselines
\begin{document}
\title{Non-minimal Wu-Yang monopole}
\author{A.B. Balakin and A.E. Zayats}
\date{ }
\maketitle
\begin{center}
{\it Department  of General Relativity and Gravitation,\\
Kazan State University, Kremlevskaya str. 18, Kazan 420008,
Russia}
\end{center}

\begin{abstract}
We discuss new exact spherically symmetric static solutions to
non-minimally extended Einstein-Yang-Mills equations. The obtained solution
to the Yang-Mills subsystem is interpreted as a non-minimal
Wu-Yang monopole solution. We focus on the analysis of two classes of the exact solutions
to the gravitational field equations. Solutions of the first class
belong to the Reissner-Nordstr{\"o}m type, i.e., they are characterized by horizons and by the
singularity at the point of origin. The solutions of
the second class are regular ones. The horizons and singularities of a new type, the non-minimal ones,
are indicated.
\end{abstract}
\noindent
PACS number(s): 04.20.Jb, 04.40.-b, 14.80.Hv\\
Keyword(s): Non-minimal Einstein-Yang-Mills theory; Exact
solutions; Wu-Yang monopole

\section{Introduction}

Exact solutions of the monopole type are known to play a
significant role in the modern field theory
\cite{Rubakov,Rossi,Milton}. The monopole solutions to the
self-consistent Einstein-Yang-Mills-Higgs equations (see, e.g.,
\cite{Bais,Cho,Yasskin,VG,Lemos06} and references therein) are of
great importance, since they demonstrate explicitly the interplay
between the gravitational, gauge and scalar fields in the
non-Abelian black hole structure formation. New possibilities for
the modeling of the monopole structure appear, when we take into
account the so-called {\it non-minimal coupling} of the
gravitational, gauge and scalar fields. Non-minimal theory has
been elaborated in detail for scalar and electromagnetic fields
(see, e.g., \cite{FaraR,HehlObukhov} for a review).
M\"uller-Hoissen obtained in \cite{MH} the non-minimal
Einstein-Yang-Mills (EYM) model from a dimensional reduction of
the Gauss-Bonnet action, this model contains one coupling
parameter. We follow the alternative derivation of the non-minimal
EYM theory, formulated as a non-Abelian generalization of the
non-minimal non-linear Einstein-Maxwell theory \cite{BL05} along
the lines proposed by Drummond and Hathrell for the linear
electrodynamics \cite{Drum}. We deal with a non-minimal EYM model
linear in curvature, which can be indicated as a
three-parameter model, since it contains three coupling constants
$q_1$, $q_2$ and $q_3$ with the dimensionality of area. Depending
on the type of the model these coupling constants may be associated with
three, two specific radii or be reduced to the unique radius, describing the
characteristic length of the non-minimal interaction (say,
$r_q=\sqrt{2|q_1|}$). Thus, in addition to the standard
Schwarzschild radius $r_g$ and Reissner-Nordstr{\"o}m radius
$r_Q$ we obtain at least one extra parameter, $r_q$, for modeling
the causal structure of non-minimally extended Einstein-Yang-Mills
monopoles.

In this letter we introduce a three-parameter self-consistent
Einstein-Yang-Mills model, in which the EYM  Lagrangian is gauge
invariant, linear in space-time curvature and quadratic in the
Yang-Mills field strength tensor ${\bf F}_{ik}$. Then we consider
exact spherically symmetric static solutions of the obtained model
and discuss in detail the non-minimal generalization of the
Wu-Yang monopole solution. We distinguish between the solutions of
the Reissner-Nordstr{\"o}m type, which is irregular in the center of the
charged body, and the regular non-minimal Wu-Yang monopole solutions. We
discuss also the relation between the values of the parameters $q_1$, $q_2$, $q_3$
and the radii of the non-minimal horizons and/or
singularities, which can be associated with the introduced new (non-minimal) Wu-Yang
monopole.

\section{ Non-minimal Einstein-Yang-Mills field equations}

The three parameter non-minimal Einstein-Yang-Mills theory can be
formulated in terms of the action functional
\begin{equation}
S_{{\rm NMEYM}} = \int d^4 x \sqrt{-g}\ \left[\frac{R}{8\pi
\gamma}+\frac{1}{2}F^{(a)}_{ik} F^{ik(a)}+\frac{1}{2}
{\cal R}^{ikmn}F^{(a)}_{ik} F^{(a)}_{mn}\right]\,.\label{act}%
\end{equation}%
Here $g = {\rm det}(g_{ik})$ is the determinant of a metric tensor
$g_{ik}$, $R$ is the Ricci scalar, $\gamma$ is the gravitational
constant. The Latin indices without parentheses run from 0 to 3,
the summation with respect to the repeated group indices $(a)$ is
implied. The tensor ${\cal R}^{ikmn}$ is defined as follows (see,
e.g., \cite{BL05,BZa06}):
\begin{equation}
{\cal R}^{ikmn} \equiv
\frac{q_1}{2}R\,(g^{im}g^{kn}-g^{in}g^{km}) +
\frac{q_2}{2}(R^{im}g^{kn} - R^{in}g^{km} + R^{kn}g^{im}
-R^{km}g^{in}) + q_3 R^{ikmn} \,, \label{sus}
\end{equation}%
where $R^{ik}$ and $R^{ikmn}$ are the Ricci and Riemann tensors,
respectively, and $q_1$, $q_2$, $q_3$ are the phenomeno\-logi\-cal
parameters describing the non-minimal coupling of the Yang-Mills
and gravitational fields. Following \cite{Rubakov} we assume that
the Yang-Mills field, ${\bf F}_{mn}$, takes the values in the Lie
algebra of the gauge group $SU(2)$:
\begin{equation}
{\bf A}_m = - i {\cal G} {\bf t}_{(a)} A^{(a)}_m  \,, \quad {\bf
F}_{mn} = - i {\cal G} {\bf t}_{(a)} F^{(a)}_{mn} \,.
\label{represent}
\end{equation}
Here ${\bf t}_{(a)}$ are  Hermitian traceless generators of
$SU(2)$ group, $A^{(a)}_i$ and $F^{(a)}_{mn}$ are the Yang-Mills
field potential and strength, respectively, the constant ${\cal
G}$ is the strength of the gauge coupling, and the group index
$(a)$ runs from $1$ to $3$. The generators ${\bf t}_{(a)}$ satisfy
the commutation relations:
\begin{equation}
\left[{\bf t}_{(a)},{\bf
t}_{(b)}\right]=i\,\varepsilon_{(a)(b)(c)}{\bf t}_{(c)}\,,
\end{equation}
where $\varepsilon_{(a)(b)(c)}$ is the completely antisymmetric
symbol with $\varepsilon_{(1)(2)(3)}=1$.

\par The variation of the action functional with respect to the
Yang-Mills potential $A^{(a)}_i$ yields
\begin{equation}
\hat{D}_k {\bf H}^{ik} \equiv \nabla_k {\bf H}^{ik}+\left[{\bf
A}_k,{\bf H}^{ik}\right] = 0 \,, \quad {\bf H}^{ik} = {\bf F}^{ik}
+ {\cal R}^{ikmn} {\bf F}_{mn} \,. \label{HikR}
\end{equation}
Here the symbol $\nabla_m$ denotes a covariant space-time derivative.
The tensor ${\bf H}^{ik}$ is a non-Abelian analogue of the induction
tensor well-known in electrodynamics \cite{Maugin}. This
analogy shows that ${\cal R}^{ikmn}$ can be considered as a
susceptibility tensor \cite{BL05}.
In a similar manner, the variation of the action with respect to
the metric yields
\begin{equation}
R_{ik} - \frac{1}{2} R \ g_{ik} = 8\pi\gamma\,T^{({\rm eff})}_{ik}
\,. \label{Ein}
\end{equation}
The effective stress-energy tensor $T^{({\rm eff})}_{ik}$ can be
partitioned into four terms:
\begin{equation}
T^{({\rm eff})}_{ik} =  T^{(YM)}_{ik} + q_1 T^{(I)}_{ik} + q_2
T^{(II)}_{ik} + q_3 T^{(III)}_{ik} \,. \label{Tdecomp}
\end{equation}
The first term $T^{(YM)}_{ik}$:
\begin{equation}
T^{(YM)}_{ik} \equiv \frac{1}{4} g_{ik} F^{(a)}_{mn}F^{mn(a)} -
F^{(a)}_{in}F_{k}^{\ n(a)} \,, \label{TYM}
\end{equation}
is a stress-energy tensor of the pure Yang-Mills field. The
definitions of other three tensors are related to the corresponding
coupling constants $q_1$, $q_2$, $q_3$:
\begin{equation}%
T^{(I)}_{ik} = R\,T^{(YM)}_{ik} -  \frac{1}{2} R_{ik}
F^{(a)}_{mn}F^{mn(a)} + \frac{1}{2} \left[ \hat{D}_{i} \hat{D}_{k}
- g_{ik} \hat{D}^l  \hat{D}_l \right] \left[F^{(a)}_{mn}F^{mn(a)}
\right] \,, \label{TI}
\end{equation}%

\[%
T^{(II)}_{ik} = -\frac{1}{2}g_{ik}\biggl[\hat{D}_{m}
\hat{D}_{l}\left(F^{mn(a)}F^{l\ (a)}_{\ n}\right)-R_{lm}F^{mn (a)}
F^{l\ (a)}_{\ n} \biggr] -{} \]%
\[{}- F^{ln(a)}
\left(R_{il}F^{(a)}_{kn} +
R_{kl}F^{(a)}_{in}\right)-R^{mn}F^{(a)}_{im} F_{kn}^{(a)} -
\frac{1}{2} \hat{D}^m \hat{D}_m \left(F^{(a)}_{in} F_{k}^{ \
n(a)}\right)+ {}\]%
\begin{equation}%
\quad{}+\frac{1}{2}\hat{D}_l \left[ \hat{D}_i \left(
F^{(a)}_{kn}F^{ln(a)} \right) + \hat{D}_k
\left(F^{(a)}_{in}F^{ln(a)} \right) \right] \,, \label{TII}
\end{equation}%

\[
T^{(III)}_{ik} = \frac{1}{4}g_{ik}
R^{mnls}F^{(a)}_{mn}F_{ls}^{(a)}- \frac{3}{4} F^{ls(a)}
\left(F_{i}^{\ n(a)} R_{knls} +
F_{k}^{\ n(a)}R_{inls}\right) -\]%
\begin{equation}%
\quad {}-\frac{1}{2}\hat{D}_{m} \hat{D}_{n} \left[ F_{i}^{ \ n
(a)}F_{k}^{ \ m(a)} + F_{k}^{ \ n(a)} F_{i}^{ \ m(a)} \right] \,.
\label{TIII}
\end{equation}%
One can check directly that the tensor $T^{({\rm eff})}_{ik}$
satisfies the equation $\nabla^k T^{({\rm eff})}_{ik} =0$, as in
the case of non-minimal electrodynamics \cite{BL05}. The self-consistent
system of equations (\ref{HikR}) and (\ref{Ein}) with
(\ref{Tdecomp}) - (\ref{TIII}) is a direct non-Abelian
generalization of the three-parameter non-minimal Einstein-Maxwell
model discussed in \cite{BL05}. This system can also be
considered as one of the variants of a non-minimal generalization of
the Einstein-Yang-Mills model.

\section{Wu-Yang monopole}

 Let us consider a static spherically symmetric non-minimal
Einstein-Yang-Mills model with the space-time metric
\begin{equation}\label{metrica}
ds^2=\sigma^2Ndt^2-\frac{dr^2}{N}-r^2 \left( d\theta^2 +
\sin^2\theta d\varphi^2 \right) \,.
\end{equation}
The Einstein-Maxwell model for such metric with a central electric
charge was studied in \cite{MHS} for the special case
$q_1+q_2+q_3=0$, $2q_1+q_2=0$. We focus on the gauge field
characterized by the special ansatz (see, \cite{RebbiRossi}):
\[%
\mathbf{A}_{0}=\mathbf{A}_{r}=0 \,,
\]%
\begin{equation}\label{1}
\mathbf{A}_{\theta}=-i\left(\frac{w}{\nu}-1\right)\mathbf{t}_{\varphi},\quad
\mathbf{A}_{\varphi}=i\left(w-\nu\right)\sin{\theta}\;\mathbf{t}_{\theta}\,.
\end{equation}
Here $\sigma$ and $N$ are functions depending on the radius
$r$ only and satisfying the asymptotic  conditions
\begin{equation}\label{asymp}
\sigma\left(\infty\right)=1 \,,\quad N\left(\infty\right)=1 \,.
\end{equation}
Generally, one can consider $w$ as a function of the radius,
however, in this paper we focus on the model with constant $w$,
keeping in mind the well-known Wu-Yang solution. The parameter
$\nu$ is a non-vanishing integer. The generators ${\bf t}_r$,
${\bf t}_{\theta}$ and ${\bf t}_{\varphi}$ are the
position-dependent ones and are connected with the standard generators of the
SU(2) group as follows:
\[%
{\bf t}_r=\cos{\nu\varphi} \ \sin{\theta}\;{\bf
t}_{(1)}+\sin{\nu\varphi} \ \sin{\theta}\;{\bf
t}_{(2)}+\cos{\theta}\;{\bf t}_{(3)},
\]%
\begin{equation}%
{\bf t}_{\theta}=\partial_{\theta}{\bf t}_r,\qquad {\bf
t}_{\varphi}=\frac {1}{\nu\sin{\theta}}\ \partial_{\varphi}{\bf
t}_r \,. \label{deS5}
\end{equation}%
The generators satisfy the relations
\begin{equation}%
\left[{\bf t}_{r},{\bf t}_{\theta}\right]=i\,{\bf t}_{\varphi}
\,,\quad \left[{\bf t}_{\theta} \,, {\bf t}_{\varphi}\right]=i\,{\bf
t}_{r} \,, \quad \left[{\bf t}_{\varphi},{\bf t}_{r}\right]=i\,{\bf
t}_{\theta}\,.\label{deS6}
\end{equation}%
The field strength tensor
\begin{equation}\label{strength}
\mathbf{F}_{ik}=\partial_i\mathbf{A}_k-\partial_k\mathbf{A}_i+\left[\mathbf{A}_i
, \mathbf{A}_k\right]
\end{equation}
has only one non-vanishing component:
\begin{equation}\label{2}
{\bf F}_{\theta\varphi}=-i\frac{(w^2-\nu^2)}{\nu}\sin\theta\,{\bf
t}_{r}\,,
\end{equation}
which does not depend on the variable $r$.
Due to the discussed ansatz the system of Yang-Mills equations
(\ref{HikR}) reduces to the single equation
\begin{equation}
\frac{w(w^2-\nu^2)}{r^4}\left(1+2{{\cal
R}^{\theta\varphi}}_{\theta\varphi}\right)=0 \,, \label{key}
\end{equation}
which is a non-minimal generalization of the well-known key
equation resulting in the Wu-Yang monopole solution
\cite{WuYang}.

There are three formal possibilities to satisfy the equation
(\ref{key}): first, $w=0$, second, $w=\pm \nu$, third,
$\left(1+2{{\cal R}^{\theta\varphi}}_{\theta\varphi}\right)=0$.
When the space-time is asymptotically flat ( $R_{ikmn}(r \to
\infty)=0$) the last term in the key equation (\ref{key}) cannot
vanish identically. When $w=\pm\nu$, we obtain from (\ref{2}) that
$\mathbf{F}_{ik}$ vanishes, and this exact solution describes the
so-called pure gauge. Finally, when $w=0$ we deal with the Wu-Yang
monopole solution. The strength of the Yang-Mills field now gets
the form $\mathbf{F}_{\theta\varphi}={i\,\nu}\sin\theta\,{\bf t}_r$, as in
the case of minimal Wu-Yang monopole in the Minkowski space-time
\cite{WuYang}. This solution is known to be effectively Abelian,
i.e., by the suitable gauge transformation ${\bf U}=\exp(-i\,{\bf
\theta \, t_\varphi})$ it can be converted into the product of the
Dirac type potential and the gauge group generator ${\bf
t}_{(3)}$.

\section{Exact solutions to the gravitational field
equations}\label{Sect}

\subsection{Key equations}

For the metric (\ref{metrica}) only four components of the
Einstein tensor $ G_i^{\,k}=R_i^{\,k}-\frac{1}{2}\delta_i^k R$ are
non-vanishing:
\begin{equation}
G_0^{\,0}=\frac{1-N}{r^2}-\frac{N'}{r} \,, \quad
G_r^{\,r}=\frac{1-N}{r^2}-\frac{N'}{r}-\frac{2N\sigma'}{r\sigma}\,,
\label{G}
\end{equation}
\begin{equation}
G_\theta^{\,\theta}=G_\varphi^{\,\varphi}=-\frac{1}{2r\sigma}\left(2\sigma
N'+2N\sigma'+3r\sigma'N'+2rN\sigma''+r\sigma N''\right).
\end{equation}
The prime denotes the derivative with respect to the radius $r$. The
corresponding four non-vanishing components of the effective
stress-energy tensor (see (\ref{TYM})-(\ref{TIII})) take the form
\begin{equation}
T_0^{\,0 ({\rm eff})}=\frac{\nu^2}{{\cal
G}^2}\left[\frac{1}{2r^4}-
q_1\frac{N'}{r^5}+(13q_1+4q_2+q_3)\frac{N}{r^6}-
(q_1+q_2+q_3)\frac{1}{r^6}\right]\,, \label{T00}
\end{equation}
\begin{equation}
T_r^{\,r ({\rm eff})}=\frac{\nu^2}{{\cal
G}^2}\left[\frac{1}{2r^4}- q_1 \frac{N'}{r^5}- 2q_1 \frac{N
\sigma'}{r^5 \sigma}-(7q_1+4q_2+q_3)\frac{N}{r^6}-
(q_1+q_2+q_3)\frac{1}{r^6}\right]\,, \label{Trr}
\end{equation}
\begin{eqnarray}
T_\theta^{\,\theta ({\rm eff})}=T_\varphi^{\,\varphi ({\rm
eff})}=-\frac{\nu^2}{{\cal
G}^2}\Biggl[\frac{1}{2r^4}-\frac{3q_1\sigma'N'}{2\sigma
r^4}-\frac{q_1N\sigma''}{\sigma
r^4}-\frac{q_1N''}{2r^4}-{}\nonumber\\{}-(7q_1+4q_2+q_3)\left(\frac{(\sigma
N)'}{\sigma r^5}-\frac{2N}{r^6}\right)+
(q_1+q_2+q_3)\frac{2}{r^6}\Biggr]\,. \label{Tpp}
\end{eqnarray}
Analogously to the case of minimal electrodynamics the equation
$G_{\theta}^{\,\theta}=8\pi\gamma\, T_\theta^{\,\theta ({\rm
eff})}$ is a differential consequence of two first Einstein
equations. Thus, in order to find two quantities, $N(r)$ and $\sigma(r)$, we have two
independent equations. Moreover, the difference of the first and
second equations,
$G_0^{\,0}-G_r^{\,r}=8\pi\gamma\left(T_0^{\,0({\rm
eff})}-T_r^{\,r({\rm eff})}\right)$, gives the equation for the
function $\sigma(r)$ only:
\begin{equation}
r \frac{\sigma^{\prime}}{\sigma} \left(1-\frac{\kappa
q_1}{r^4}\right)= \frac{\kappa (10q_1+4q_2+q_3)}{r^4}\,.
\label{si}
\end{equation}
Here $\kappa=\frac{8\pi\gamma\nu^2}{{\cal G}^2}$ is a new
convenient constant with the dimensionality of area. The function
$N(r)$ satisfies the linear differential equation
\begin{equation}
r N^{\prime} \left(1-\frac{\kappa q_1}{r^4}\right) + N \left[1 +
\frac{\kappa}{r^4} (13q_1 +4q_2 +q_3) \right] = 1 -
\frac{\kappa}{2r^2} + \frac{\kappa}{r^4} (q_1 +q_2 +q_3) \,.
\label{N}
\end{equation}
It is worth mentioning that all the non-minimal contributions to the equations
(\ref{si}) and (\ref{N}) have the similar form: they contain
products of $\kappa$ and linear combinations of the coupling
constants divided by $r^4$.

\subsection{Minimal limit $q_1=q_2=q_3=0$}

When $q_1$, $q_2$, $q_3$ vanish, the equation (\ref{si}) and the
asymptotic conditions (\ref{asymp}) give $\sigma(r)=1$. The
equation (\ref{N}) yields
\begin{equation}\label{RN}
N = 1 - \frac{2M}{r} + \frac{\kappa}{2r^2}\,,
\end{equation}
where $M$ is a constant of integration describing the asymptotic
mass of the monopole (in the geometrical units $2M$ is equal to
the Schwarzschild radius $r_g$). This solution is of the
Reissner-Nordstr{\"o}m type.

\subsection{Non-minimal models with $q_1 \neq 0$}

For generic $q_1$, $q_2$, $q_3$ the equations (\ref{si}) and
(\ref{N}) with the conditions (\ref{asymp}) yield
\begin{equation}\label{si2}
\sigma=\left(1-\frac{\kappa q_1}{r^4}\right)^{\beta}\,, \quad \beta
\equiv \frac{10q_1+4q_2+q_3}{4q_1} \,,
\end{equation}
\begin{equation}
N=1-\frac{1}{r}\cdot\left(1-\frac{\kappa
q_1}{r^4}\right)^{-(\beta+1)} \left\{2M-\frac{\kappa}{2}
\int\limits_r^{+\infty}\frac{dx}{x^2}\left[1+\frac{6}{x^2}(4q_1+q_2)\right]\left(1-\frac{\kappa
q_1}{x^4}\right)^{\beta}\right\}\,.\label{N2}
\end{equation}
When $r \to \infty$, these solutions asymptotically behave as
\begin{equation}
\sigma=1 - \frac{\kappa q_1}{r^4}\,\beta + \dots\,,\quad
N=1-\frac{2M}{r}+\frac{\kappa}{2r^2}+ \frac{\kappa}{r^4}\,
(4q_1+q_2)+ \dots \,,
\end{equation}
thus the leading order terms recover the Reissner-Nordstr{\"o}m
solution, and the non-minimal contributions contain the terms
$\frac{1}{r^4}$, $\frac{1}{r^5}$, etc.

When $r \to 0$, the function $\sigma(r)$ can tend to infinity, if
$\beta$ is positive, can tend to zero, if $\beta$ is negative, and
remains equal to one, if $\beta=0$. In other words, the metric
coefficient $g_{00}=\sigma^2 N$ can be irregular at the point of origin
$r=0$, when $10q_1+4q_2+q_3\neq 0$. Moreover, when the
parameter $q_1$ is positive, the metric coefficient $\sigma(r)$
takes zero value at the point $r_{({\rm s})} = (\kappa
q_1)^{\frac{1}{4}}$, if $\beta > 0$, and becomes infinite, if
$\beta <0$, providing the curvature invariants to be infinite
at $r_{\rm (s)}$. To illustrate this remark, let us assume that
$\beta=1$, or, equivalently, $6q_1+4q_2+q_3=0$. Then one obtains
the exact solution
\begin{equation}\label{si22}
\sigma=1-\left(\frac{r_{({\rm s})}}{r}\right)^4\,,
\end{equation}
\begin{equation}
N=1- \left[1- \left(\frac{r_{({\rm
s})}}{r}\right)^4\right]^{-2}\left\{ \frac{2M}{r} -
\frac{\kappa}{2r^2}\left[1 - \frac{1}{5}\left(\frac{r_{({\rm
s})}}{r}\right)^4  +\frac{2(4q_1+q_2)}{r^2}\left[1- \frac{3}{7}
\left(\frac{r_{({\rm s})}}{r}\right)^4 \right]\right]\right\}
\,,\label{N22}
\end{equation}
for which $\sigma(r_{({\rm s})})=0$. As for $N(r_{({\rm s})})$, it
can be infinite, equal to zero or take a finite value depending
on relationships between $q_1$, $q_2$, $M$ and $\kappa$. For
instance, when $q_1=\frac{\kappa}{16}$, $q_2=-\frac{\kappa}{8}$ and
$M=\frac{24}{35}\sqrt{\kappa}$ one obtains that $r_{({\rm
s})}=\frac{\sqrt{\kappa}}{2}$ and $N(r_{({\rm s})})=0$, however, the
Ricci scalar $R$ and quadratic curvature invariants $R_{ik}R^{ik}$,
$R_{ikmn}R^{ikmn}$ are regular at the point $r_{({\rm s})}$. In
other cases these invariants become infinite, and the point $r=r_{({\rm s})}$
can be indicated as a specific non-minimal singularity.
When $q_1$ is negative, such a singularity does not
appear.

\subsection{Non-minimal models with $q_1 = 0$}

Since $q_1$ appears in the denominator of the expression
(\ref{si2}) for $\beta$, let us consider the case $q_1 = 0$ as a special one.
Now the metric functions are
\begin{equation}\label{si1}
\sigma = \exp\left\{-\frac{\kappa}{4r^4}(4q_2+q_3) \right\}\,,
\end{equation}
\begin{equation}
N=1-\frac{1}{r}\cdot\exp\left\{{\frac{\kappa(4q_2+q_3)}{4r^4}}\right\}\cdot\left(2M-\frac{\kappa}{2}
\int\limits_r^{+\infty}\frac{dx}{x^2}\left(1+\frac{6q_2}{x^2}\right)\exp\left\{{-\frac{\kappa(4q_2+q_3)}
{4x^4}}\right\}\right)\,.\label{N1}
\end{equation}
Clearly, the analytical progress is possible, when
$q_3=-4q_2$. Indeed, for this model $\sigma(r)=1$, and the explicit
exact solution for the function $N(r)$ is
\begin{equation}
N=1-\frac{2M}{r} + \frac{\kappa}{2r^2} + \frac{\kappa q_2}{r^4}
\,.\label{N11}
\end{equation}
We deal with the non-minimal generalization of the
Reissner-Nordstr{\"o}m star with $N(0) = \infty$. Such a star
possesses horizons, when the algebraic equation of the fourth
order
\begin{equation}
r^4 - 2M r^3 + \frac{\kappa}{2}r^2 + \kappa q_2=0 \label{N111}
\end{equation}
has real positive roots. There are two explicit cases admitting
specific non-minimal horizons.

\noindent
{\it (i) $M=0  \ \  {\rm and }  \ \ q_2<0$}

\noindent
Then the positive real root of (\ref{N111}) is
\begin{equation}
r= r_{({\rm H})}=\frac{1}{2} \sqrt{\kappa}\sqrt{\sqrt{1+\frac{16
|q_2|}{\kappa}}-1} \,. \label{N112}
\end{equation}
In the minimal limit $r_{({\rm H})}$ coincides with $r=0$ and tends
to $\sqrt{2|q_2|}$ when $|q_2| \ll \kappa$.

\noindent
{\it (ii) $\kappa = 2M^2 \ \ {\rm  and } \ \ q_2<0$}

\noindent
The equation (\ref{N111}) possesses the positive real root
\begin{equation}
r_{({\rm H}1)}= \frac{M}{2}\left(1+\sqrt{1+\frac{4
\sqrt{2|q_2|}}{M}} \right) \,. \label{N113}
\end{equation}
When $M>4\sqrt{2|q_2|}$, there are two additional roots
\begin{equation}
r_{({\rm H}2,3)}= \frac{M}{2}\left(1 \pm \sqrt{1-\frac{4
\sqrt{2|q_2|}}{M}} \right)\,. \label{N114}
\end{equation}
In the minimal limit the condition $\kappa = 2M^2$
leads to the so-called extremal Reissner-Nordstr{\"o}m black
hole, for which two horizons coincide. When $q_2<0$, the specific
radii $r_{({\rm H}1)}$, $r_{({\rm H}2)}$ and $r_{({\rm H}3)}$ play
the roles of the non-minimal horizons radii. When
$q_2$ tends to zero, $r_{({\rm H}1)} \to r_{({\rm H}2)} \to M$ and
$r_{({\rm H}3)} \to 0$.

\subsection{Regular one-parameter model}

\par When $10q_1+4q_2+q_3=0$, $4q_1+q_2=0$, i.e., $q_1=-q$, $q_2=4q$, $q_3=-6q$, and $q$ is positive,
 we obtain a new explicit exact solution
\begin{equation}\label{sN1}
\sigma(r)=1\,,\quad N=1+\frac{r^2\,(k-4Mr)}{2\,(r^4+\kappa q)} \,.
\end{equation}
The obtained function $N(r)$ takes the value $N=1$ at three
points: $N(0)=1$, $N \left(\frac{\kappa}{4M}\right)=1$,
$N(\infty)=1$ (asymptotically). When $M=0$ the second and the
third points coincide, $N(r)\geq 1$ and $N(r)$ has only one
extremum (maximum) at the point $r_{({\rm max})} = (\kappa
q)^{\frac{1}{4}}$. For small $M$ one has a
minimum at some point $r_{({\rm min})}$ ($r_{({\rm min})} >
\frac{\kappa}{4M}$), for which $0 < N_{({\rm min})} < 1$. When the mass $M$ increases,
this minimum reaches the value $N_{({\rm min})}=0$ with the mass
taking a critical value $M_{({\rm crit})}$ of the following form
\begin{equation}
M_{{(\rm
crit})}=\frac{r_{\!*}}{6}\left(4+\frac{\kappa}{r_{\!*}^2}\right)\,.
\label{Mcrit}
\end{equation}
Here
\begin{equation}
r_{\!*}=\frac{\sqrt{\kappa}}{2}\sqrt{\left(\sqrt{1+\frac{48\,q}{\kappa}}+1\right)}\,.
\label{rstar}
\end{equation}
Thus, when $M<M_{({\rm crit})}$ the metric (\ref{sN1}) has no
horizons; when $M>M_{({\rm crit})}$ there are two horizons,
$r_{-}$ and $r_{+}$. When $M=M_{({\rm crit})}$ the function $N(r)$
takes zero value  only at $r=r_{\!*}$, i.e., in this case the
metric (\ref{sN1}) is a non-minimal analogue of the extremal
Reissner-Nordstr{\"o}m solution. When $q=0$, the parameter
$r_{\!*}$ coincides with the Reissner-Nordstr{\"o}m radius,
$r_Q=\sqrt{\frac{\kappa}{2}}$.

The solution (\ref{sN1}) is regular at the point $r=0$, since the
denominator cannot reach zero value. In addition, direct
calculations show that the curvature invariants $R$,
$R_{ik}R^{ik}$, $R_{ikmn}R^{ikmn}$ take finite values at $r=0$.

\section{Conclusions}

We have shown that the three-parameter non-minimally extended
Einstein-Yang-Mills theory admits the exactly
solvable generalization of the Wu-Yang monopole model. Indeed,
the non-minimal Yang-Mills subsystem admits the
exact solution of the standard explicit form (\ref{1}),
(\ref{2}). The solutions to the gravitational field equations are also
presented in the explicit form: in the quadratures for generic $q_1$, $q_3$ and $q_3$
(see, (\ref{N2}), (\ref{N1})), and in the elementary
functions for the special choices of the coupling parameters (see,
(\ref{N22}), (\ref{N11})). The analysis of these exact solutions
permits the following three features to be emphasized.

\vspace{3mm}
\noindent
{\it (i) On the inheritance of the structure of the Wu-Yang monopole solution.}

\noindent Non-minimal interaction of the Yang-Mills and
gravitational fields results in essentially complicated master
equations (see, (\ref{HikR}), (\ref{sus})). Nevertheless, the
well-known Wu-Yang solution with the ansatz (\ref{1}) keeps its form
in the non-minimally extended theory, the coupling parameters $q_1$,
$q_2$ and $q_3$ do not enter the expression for ${\bf A}_i$.

\vspace{3mm}
\noindent
{\it (ii) On the regularity of the Wu-Yang monopole.}

\noindent The analytical solution (\ref{sN1}) to the gravitational field equations
is regular at $r=0$ ($\sigma(0)=1$,
$N(0)=1$) and has no horizons, when $M<M_{({\rm crit})}$. The
curvature invariants, $R$, $R_{ik}R^{ik}$, $R_{ikmn}R^{ikmn}$, for
such a gravity field are finite for arbitrary $r$. In contrast to
the curvature invariants the invariant of the gauge field,
$I_{(1)} = \frac{1}{2}F^{(a)}_{ik} F^{(a) ik}= \frac{\nu^2}{{\cal
G}^2 r^4}$, is singular at $r=0$. Thus, we give an example, which demonstrates
that the non-minimal interaction can eliminate the
singularity of the gravitational field.

\vspace{3mm} \noindent {\it (iii) On the non-minimal horizons and
singularities.}

\noindent The formulas in Sec.\ref{Sect} show that the space-time
metric, describing the gravitational field of the Wu-Yang monopole,
can contain a number of horizons and singularities depending on
the relationships between $q_1$, $q_2$, $q_3$, as well as on their signs
and values. When the coupling constants vanish, all these horizons
and singularities convert into inner, outer
horizons and point of origin for the Reissner-Nordstr{\"o}m
metric, respectively. In other words, the non-minimal coupling splits the
characteristic surfaces, and makes the causal structure of the
object much more sophisticated. This problem requires a special
discussion.

\section*{Acknowledgements}
This work was partially supported by the Deutsche
Forschungsgemeinschaft. A.B. is grateful to Professors H.~Dehnen,
W.~Zimdahl and J.P.S.~Lemos for stimulating discussions.

\end{document}